\documentclass[cits]{PoS}

\usepackage{enumitem}
\usepackage{amsmath}
\usepackage{xspace}
\usepackage{slashed}
\usepackage{cite}
\usepackage{bm}

\def\refeq#1{\mbox{(\ref{#1})}}
\def\reffi#1{\mbox{Figure~\ref{#1}}}

\def\reffi#1{\mbox{Fig.~\ref{#1}}}

\def\citere#1{\mbox{Ref.~\cite{#1}}}
\def\citeres#1{\mbox{Refs.~\cite{#1}}}

\newcommand{\cmm}[1]{\ensuremath{#1}\ifmmode\else{}\fi}
\newcommand{\nmc}[2]{\newcommand{#1}{\cmm{#2}}}
\nmc{\al}{\alpha}
\nmc{\be}{\beta}
\nmc{\de}{\delta}
\nmc{\la}{\lambda}
\nmc{\si}{\sigma}

\newcommand{\ri}{\mathrm i}
\newcommand{\ie}{{i.e.}\ }
\newcommand{\eg}{{e.g.}\ }

\nmc{\rd}{\mathrm{d}}

\def\beq{\begin{equation}}
\def\eeq{\end{equation}}

\newcommand{\PH}{\ensuremath{\text{H}}\xspace}
\newcommand{\PHone}{\ensuremath{{\text{H}_1}}\xspace}
\newcommand{\PHtwo}{\ensuremath{{\text{H}_2}}\xspace}
\newcommand{\PHpm}{\ensuremath{\text{H}^\pm}\xspace}
\newcommand{\PHa}{\ensuremath{A_0}\xspace}

\newcommand{\PW}{\ensuremath{\text{W}}\xspace}
\newcommand{\PZ}{\ensuremath{\text{Z}}\xspace}

\newcommand{\MW}{\ensuremath{M_\PW}\xspace}

\newcommand{\GeV}{\ensuremath{\,\text{GeV}}\xspace}

\newcommand{\UV}{{\mathrm{UV}}}
\newcommand{\rw}{{\mathrm{w}}}
\nmc{\alem}{\alpha_{\mathrm{em}}}
\nmc{\sw}{s_{\rw}}
\nmc{\cw}{c_{\rw}}

\nmc{\g}{g_2}
\nmc{\gy}{g_1}
\nmc{\Gf}{G_\mathrm{F}}

\newcommand{\cL}{\ensuremath{\mathcal{L}}\xspace}
\newcommand{\cM}{\ensuremath{\mathcal{M}}\xspace}
\newcommand{\rT}{\ensuremath{\text{T}}\xspace}
\newcommand{\rL}{\ensuremath{\text{L}}\xspace}
\newcommand{\rR}{\ensuremath{\text{R}}\xspace}

\newcommand{\hc}{\ensuremath{\text{h.c.}}\xspace}
\nmc{\rB}{{\rm B}}
\nmc{\rs}{{\rm s}}

\nmc{\Msb}{M_{\rm sb}}

\nmc{\ftone}{t_{s}}
\nmc{\tab}{t_{\alpha\beta}}

\nmc{\tHoneHone}{t_{H_1H_1}}
\nmc{\tHoneHtwo}{t_{H_1H_2}}
\nmc{\tHtwoHtwo}{t_{H_2H_2}}
\nmc{\tHaHa}{t_{\Ha\Ha}}
\nmc{\tHpmHpm}{t_{H^\pm H^\pm}}
\nmc{\tGzGz}{t_{G_0G_0}}
\nmc{\tGpmGpm}{t_{G^\pm G^\pm}}
\nmc{\tGzHa}{t_{G_0\Ha}}
\nmc{\tGpmHpm}{t_{G^\pm H^\pm}}

\newcommand{\nunuone}{{\nu_1}\bar\nu_1}
\newcommand{\nunutwo}{{\nu_2}\bar\nu_2}

\newcommand{\Prophecy}{{\sc Prophecy4f}}

\def\draftdate{\relax}
\def\mda{\relax}
\def\mua{\relax}
\def\mla{\relax}
\def\draft{
\def\thtystars{******************************}
\def\sixtystars{\thtystars\thtystars}
\typeout{}
\typeout{\sixtystars**}
\typeout{* Draft mode!
         For final version remove \protect\draft\space in source file *}
\typeout{\sixtystars**}
\typeout{}
\def\draftdate{\today}
\def\mua{\marginpar[\boldmath\hfil$\uparrow$]%
                   {\boldmath$\uparrow$\hfil}%
                    \typeout{marginpar: $\uparrow$}\ignorespaces}
\def\mda{\marginpar[\boldmath\hfil$\downarrow$]%
                   {\boldmath$\downarrow$\hfil}%
                    \typeout{marginpar: $\downarrow$}\ignorespaces}
\def\mla{\marginpar[\boldmath\hfil$\rightarrow$]%
                   {\boldmath$\leftarrow $\hfil}%
                    \typeout{marginpar: $\leftrightarrow$}\ignorespaces}
\def\Mua{\marginpar[\boldmath\hfil$\Uparrow$]%
                   {\boldmath$\Uparrow$\hfil}%
                    \typeout{marginpar: $\uparrow$}\ignorespaces}
\def\Mda{\marginpar[\boldmath\hfil$\Downarrow$]%
                   {\boldmath$\Downarrow$\hfil}%
                    \typeout{marginpar: $\downarrow$}\ignorespaces}
\def\Mla{\marginpar[\boldmath\hfil{$\Rightarrow$}]%
                   {\boldmath{$\Leftarrow $}\hfil}%
                    \typeout{marginpar:$\leftrightarrow$}\ignorespaces}
\def\muanick{\marginpar[\boldmath\hfil$\uparrow$]%
                   {\boldmath$\textcolor{blue}\uparrow$\hfil}%
                    \typeout{marginpar:\textcolor{blue} $\uparrow$}\ignorespaces}
\def\mdanick{\marginpar[\boldmath\hfil$\downarrow$]%
                   {\boldmath$\textcolor{blue}\downarrow$\hfil}%
                    \typeout{marginpar: $\downarrow$}\ignorespaces}
\def\mlanick{\marginpar[\boldmath\hfil$\rightarrow$]%
                   {\boldmath$\textcolor{blue}\leftarrow $\hfil}%
                    \typeout{marginpar: $\leftrightarrow$}\ignorespaces}
\overfullrule 5pt
\oddsidemargin 15mm
\marginparwidth 29mm
}

\nmc{\Hhat}{\hat H}
\nmc{\hhat}{\hat h}
\nmc{\Phihat}{\hat \Phi}
\nmc{\phihat}{\hat \phi}
\nmc{\chihat}{\hat \chi}
\nmc{\etahat}{\hat \eta}
\nmc{\rhohat}{\hat \rho}
\nmc{\thetahat}{\hat \theta}
\nmc{\sihat}{\hat\sigma}
\nmc{\Phione}{\Phi_1}
\nmc{\Phitwo}{\Phi_2}
\nmc{\Zhat}{\hat Z}
\nmc{\Hahat}{\hat A_0}
\nmc{\Gzhat}{\hat G_0}
\nmc{\Xhat}{\hat X}
\nmc{\Yhat}{\hat Y}
\nmc{\Honehat}{\hat H_1}
\nmc{\Htwohat}{\hat H_2}

\nmc{\vone}{v_1}
\nmc{\vtwo}{v_2}
\nmc{\etaone}{\eta_1}
\nmc{\etatwo}{\eta_2}
\nmc{\etaplet}{\bm{\eta}}
\nmc{\chione}{\chi_1}
\nmc{\chitwo}{\chi_2}
\nmc{\chiplet}{\boldsymbol{\chi}}
\nmc{\Hone}{H_1}
\nmc{\Htwo}{H_2}
\nmc{\Hplet}{\mathbf{H}}
\nmc{\phionepm}{\phi_1^\pm}
\nmc{\phitwopm}{\phi_2^\pm}
\nmc{\Ha}{A_0}

\nmc{\tb}{t_\be}
\nmc{\ca}{c_\al}
\nmc{\catwo}{c^2_\al}
\nmc{\satwo}{s^2_\al}
\nmc{\sa}{s_\al}
\nmc{\stwoa}{s_\al}
\nmc{\cbe}{c_\be}
\nmc{\ctwobe}{c_{2\be}}
\nmc{\cbetwo}{c^2_\be}
\nmc{\sbe}{s_\be}
\nmc{\sbetwo}{s^2_\be}
\nmc{\cab}{c_{\al\be}}

\nmc{\dth}{\delta t_{H}}
\nmc{\dthhat}{\delta t_{\hat{H}}}
\nmc{\Th}{T^{H}}
\nmc{\Thhat}{T^{\Hhat}}
\nmc{\THone}{{T}^{H_1}}
\nmc{\THtwo}{{T}^{H_2}}
\nmc{\dtHone}{{\delta t}_{H_1}}
\nmc{\dtHtwo}{{\delta t}_{H_2}}
\nmc{\dtHhatone}{{\delta t}_{\hat H_1}}
\nmc{\dtHhattwo}{{\delta t}_{\hat H_2}}
\nmc{\MHone}{{M}_\PHone}
\nmc{\MHtwo}{{M}_\PHtwo}
\nmc{\MHa}{{M}_{\PHa}}
\nmc{\MHpm}{{M}_{\text{H}^\pm}}
\nmc{\MHsone}{{M}^2_\PHone}
\nmc{\MHstwo}{{M}^2_\PHtwo}
\nmc{\MHsa}{{M}^2_{\PHa}}
\nmc{\MHspm}{{M}^2_{\PHpm}}

\newcommand{\THDM}{THDM\xspace}

\newcommand{\MSbar}{\ensuremath{\overline{\text{MS}}}\xspace}

\newcommand{\OSonetwo}{OS12\xspace}

\nmc{\vshift}{\bar{v}}

\newcommand{\U}{\mathrm{U}}
\newcommand{\SU}{\mathrm{SU}}

\title{Renormalization schemes for mixing angles \\
{in extended Higgs sectors}}

\ShortTitle{Renormalization schemes for mixing angles}

\author{\speaker{Ansgar Denner}%\thanks{A footnote may follow.}
        \\
        Universit\"at W\"urzburg, %
        Institut f\"ur Theoretische Physik und Astrophysik, \\  %
        97074 W\"urzburg, %
        Germany\\
        E-mail: \email{Ansgar.Denner@uni-wuerzburg.de}}

\author{Stefan Dittmaier\\
        Albert-Ludwigs-Universit\"at Freiburg, %
        Physikalisches Institut, %
        79104 Freiburg, %
        Germany\\
        E-mail: \email{stefan.dittmaier@physik.uni-freiburg.de}}

\author{Jean-Nicolas Lang\\
        Universit\"at Z\"urich, 
        Physik-Institut, 
        CH-8057 Z\"urich,
        Switzerland\\
        E-mail: \email{jlang@physik.uzh.ch}}
      
      \abstract{The proper renormalization of mixing angles in quantum
        field theories is a long-standing problem. It is relevant for
        the renormalization of the quark mixing matrix in the Standard
        Model and for various mixing scenarios in theories beyond. In
        this contribution we specifically consider theories with
        extended scalar sectors. We describe renormalization schemes
        for mixing angles based on combinations of observables or
        symmetry requirements such as rigid or background-field gauge
        invariance {and compare their properties to previous approaches
        such as $\MSbar$ schemes.}
        We {formulate} specific renormalization conditions
        for the mixing angles in the Two-Higgs-Doublet Model and the
        Higgs-Singlet Extension of the Standard Model and calculate
        electroweak corrections {to Higgs-boson decays via W- or
        Z-boson pairs} within these models for a selection of
        {(new and old)} renormalization schemes.}

\FullConference{14th International Symposium on Radiative Corrections (RADCOR2019)\\ 
                9-13 September 2019\\
                Palais des Papes, Avignon, France}

\begin{document}

\section{Introduction}

Mixing of different fields with identical conserved quantum numbers
appears in many theories. The Standard Model contains mixing angles in
the fermion sector, the parameters of the quark-mixing matrix. In
models with extended Higgs sectors, different scalar fields typically
mix, while in models with extended gauge sectors, mixing between
different gauge fields takes place. The investigation of the Higgs
sector is presently in the focus of particle physics {at the LHC.}  
This requires
to perform precise calculations in models with extended Higgs sectors
and in turn the renormalization of these models including the
renormalization of mixing angles.  In this contribution we focus on
the renormalization of mixing angles in extended Higgs sectors.

Renormalization of mixing angles in extended Higgs sectors has been
performed in various models. The renormalization of the mixing angle
$\beta$ in the Minimal Supersymmetric Standard Model (MSSM) was
investigated in
\citeres{Chankowski:1992er,Dabelstein:1994hb,Freitas:2002um,Baro:2008bg,%
Baro:2009gn}.
Renormalization conditions for the mixing angles $\alpha$ and $\beta$
of the Two-Higgs-Doublet Model (THDM) were discussed in
\citeres{Kanemura:2004mg,LopezVal:2009qy,Kanemura:2014dja,%
Kanemura:2015mxa,Krause:2016oke,Denner:2016etu,Denner:2017vms,%
Altenkamp:2017ldc}, while the mixing-angle renormalization for the
Higgs-Singlet Extension of the Standard Model (HSESM) was investigated
in
\citeres{ Denner:2017vms,Kanemura:2015fra,Bojarski:2015kra,Kanemura:2016lkz,%
 Altenkamp:2018bcs}. Finally, the renormalization of
mixing angles in the Next-to-Two-Higgs-Doublet Model was performed in
\citere{Krause:2017mal}.

Previous {work} on the renormalization of mixing angles {has revealed}
problems in various schemes. $\MSbar$ renormalization of mixing angles
can lead to perturbative instabilities. Fixing the renormalized mixing
angles from individual physical observables violates symmetries and
can give rise to unnaturally large corrections, while
renormalization conditions relying on self-energies instead of
physical quantities introduce gauge dependences.

Based on the study of the renormalization of $\tan\beta$ in the MSSM,
Freitas and St\"ockinger \cite{Freitas:2002um} formulated desirable
properties for the renormalization of mixing angles. Accordingly,
mixing-angle renormalization should be (the last condition was added
in {\citere{Denner:2018opp}):}
\begin{itemize}\itemsep -3pt
\item {\em gauge independent}, \ie renormalized observables should be
  gauge-independent functions of the renormalized mixing angles;
\item {\em symmetric} with respect to the mixing degrees of freedom,
  \ie in particular independent of a specific physical process;
\item {\em perturbatively stable}, \ie higher-order corrections should
  not be artificially large;
\item {\em non-singular} for degenerate masses of mixing fields and
  for extreme mixing angles, \ie valid in the full parameter space.
\end{itemize}
The aim of this contribution is to {recapitulate the renormalization
conditions for mixing angles introduced in \citere{Denner:2018opp} that 
fulfil these conditions.
For more details and further phenomenological applications}
we refer to \citere{Denner:2018opp}.

\section{Tadpoles and renormalization}

Field theories with spontaneous symmetry breaking involve scalar
fields with non-vanishing vacuum expectation (vevs) values
$\langle\Phi\rangle=v\ne0$. Since perturbation theory requires an
expansion around the minimum of the potential, shifted fields with
vevs are introduced, $\Phi(x)=\bar{v}+H(x)$, with $\langle H\rangle=0$
at least in leading order (LO). If the expansion is performed about
the LO vev $v_0$, at next-to-leading order (NLO) so-called tadpole
contributions $T^H$ from one-particle irreducible diagrams with a
single external Higgs field appear. These can be eliminated via an
expansion about the corrected vev $v=v_0+\Delta v$, which leads to
tadpole counterterms $\de t^H$ that cancel the tadpole contributions
of loop diagrams.

In the conventional tadpole scheme, which has been used in the
Standard Model for instance in
\citeres{tHooft:1971qjg,Passarino:1978jh,Aoki:1982ed,Denner:1991kt,Actis:2006},
the corrected vev $v$ is used throughout. As a consequence, no
explicit tadpole diagrams appear, and the Ward identities are
simplified owing to $\langle H\rangle=0$ in all orders. However, the
corrected vev is related to the tadpoles, which are gauge dependent.
As a consequence, the bare masses of the gauge bosons, fermions, and
the Higgs boson as well as the corresponding counterterms become gauge
dependent.  This does not matter if all observables are expressed in
terms of observables such as physical masses, which is the case in the
on-shell schemes used for the Standard Model.  If, however, some
parameters are renormalized in an $\MSbar$ scheme, these parameters
and the {$S$-matrix} as a function thereof become gauge dependent. This
is, in particular, the case for mixing angles in extended scalar
sectors renormalized within the $\MSbar$ scheme
\cite{Krause:2016oke,Denner:2016etu}. While the conventional tadpole
scheme appears in the literature in variants differing in the
definition of the Higgs-boson mass, all variants share {the 
properties} discussed above. We use in the following the variant
defined in {\citeres{Denner:2016etu,Denner:1991kt}} dubbed PRTS in
\citere{Denner:2018opp}.

An alternative tadpole scheme was introduced in
\citere{Fleischer:1980ub}, dubbed FJTS. It uses consistently the bare
(LO) vev $v_0$. As a consequence, all renormalized parameters are
gauge independent also in $\MSbar$ schemes.  On the down side,
$\langle H\rangle\ne0$ beyond LO, and tadpoles appear in many places,
\eg in the definition of the renormalized masses and in Ward
identities. Upon performing {the shift $H\to H+\Delta v$ of the Higgs
  field~$H$ by the constant $\Delta v$,} the explicit tadpoles can be
replaced by contributions of $\Delta v$ and thus transferred to the
counterterms of vertex functions with more than one external leg.
{Even though being fully consistent, the FJTS is prone to large
  corrections originating from $\MSbar$-renormalized mixing
  angles~\cite{Altenkamp:2017ldc,Altenkamp:2017kxk,Altenkamp:2018bcs,Denner:2018opp}.}

\section{Extended Higgs {sectors}}

The mixing between two CP-even scalars appears in many models with
extended Higgs sectors. Denoting the fields in the symmetric basis by
$\eta_1$, $\eta_2$ and the fields in the physical mass-eigenstate
basis by $H_1$, $H_2$, the mixing is described by
\begin{align}
  \etaplet = \left(\begin{array}{c}
    \etaone\\
    \etatwo
  \end{array}\right)=
  R(\al)
  \left(\begin{array}{c}
    \Hone\\
    \Htwo
  \end{array}\right) =  R(\al) \Hplet, \qquad
  R(\alpha)=
  \left(\begin{array}{cc}
        \ca & -\sa\\
        \sa & ~\ca
  \end{array}\right),
  \label{eq:scalar_rotation}
\end{align}
with the shorthand notations $\ca=\cos\alpha$ and $\sa=\sin\al$ for
the mixing angle $\al$.

The renormalization transformations for $\al$ %the mixing angle 
and the
scalar fields in the physical basis read
\begin{equation}\label{eq:cms_renormalization}
\al_{\rB} = \al+\de\al,\qquad
\Hplet_{\rB} ={} (Z^H)^{1/2} \Hplet,
\end{equation}
where the index $\rB$ denotes bare quantities and 
the field renormalization constants are parametrized as
\begin{equation}
{  (Z^H)^{1/2}={} }
%  \left(\begin{array}{cc}
%        (Z^H)_{11}^{1/2} &  (Z^H)_{12}^{1/2}\\
%        (Z^H)_{21}^{1/2} &  (Z^H)_{22}^{1/2}
%  \end{array}\right)
%= {\bf 1} + \frac{1}{2}\de Z^H
\left(\begin{array}{cc}
        1 + \frac{1}{2}\de Z^H_{11} &   \frac{1}{2}\de Z^H_{12}\\
         \frac{1}{2}\de Z^H_{21} &  1+\frac{1}{2}\de Z^H_{22}
  \end{array}\right).
\label{eq:matrix_renormalization_scalars}
\end{equation}

In the complete on-shell scheme \cite{Aoki:1982ed,Denner:1991kt} the
non-diagonal field renormalization constants are fixed as
\beq\label{eq:zij_onshell}
\delta Z^H_{ij} = \frac{2}{M_{\PH_i}^2-M_{\PH_j}^2}\Sigma_{ij}(M_{\PH_j}^2), \quad i\ne j,
\eeq
where $M_{\PH_i}$ and $M_{\PH_j}$ are the masses of the
scalar bosons $\PH_i$ and $\PH_j$, respectively, and $\Sigma_{ij}$
their mixing energy.  

\section{Renormalization schemes for mixing angles based on symmetries}

Symmetric renormalization conditions for mixing angles can be obtained
upon using rigid symmetry, \ie the symmetry under global $\SU(2)\times
\U(1)$ transformations. Since a spontaneously broken theory can be
renormalized in the unbroken phase
%\cite{tHooft:1971qjg,tHooft:1971akt,%
%Lee:1972fj,Lee:1974zg,Lee:1972yfa,Lee:1973fn,Lee:1973rb},
\cite{tHooft:1971qjg,Lee:1974zg,Lee:1973rb},
the following renormalization transformations are sufficient to absorb
the ultraviolet (UV) singularities
\begin{align}
\al_{\rB} ={}& \al+\de\al,\qquad
{\etaplet}_{\rB} ={} (Z^{\eta})^{1/2} {\etaplet}, \\
  (Z^\eta)^{1/2}={}&
  \left(\begin{array}{cc}
        (Z^{\eta}_{1})^{1/2} &  0\\
        0 &  (Z^{\eta}_2)^{1/2}
  \end{array}\right)
%=1+\frac{1}{2}\delta Z^{\eta}
= \left(\begin{array}{cc}
        1 + \frac{1}{2}\de Z^{\eta}_1 &  0\\
        0 &  1+\frac{1}{2}\de Z^{\eta}_2
  \end{array}\right).
\end{align}
Consistency with renormalization in the complete on-shell scheme
\refeq{eq:cms_renormalization},
\refeq{eq:matrix_renormalization_scalars} requires for the
UV-divergent parts
\begin{align}
  (Z^H)^{1/2}\big|_\UV = R^{\rT}(\al+\de\al) (Z^{\eta})^{1/2}
  R(\al)\big|_\UV
\end{align}
and as a consequence
\begin{align}
\label{eq:Zsymrel}
\de Z^H_{11}\big|_\UV ={}& \ca^2\de Z^{\eta}_1\big|_\UV + \sa^2\de Z^{\eta}_2\big|_\UV, \notag\\
\de Z^H_{22}\big|_\UV ={}& \sa^2\de Z^{\eta}_1\big|_\UV + \ca^2\de Z^{\eta}_2\big|_\UV, \notag\\
\de Z^H_{12}\big|_\UV+\de Z^H_{21}\big|_\UV ={}& 2\ca\sa(\de Z^{\eta}_2-\de Z^{\eta}_1)\big|_\UV, \notag\\
\de Z^H_{12}\big|_\UV-\de Z^H_{21}\big|_\UV ={}& 4\de\al\big|_\UV,
\end{align}
\ie UV divergences of mixing-angle counterterms are related to UV
divergences of $\de Z^H_{ij}$.  This can be used to fix the
renormalization of $\alpha$ via
\cite{Kanemura:2004mg,Krause:2016oke,Denner:2018opp}
\begin{align}\label{de_al_rigid}
\de\al = \frac{1}{4}\bigl(\de Z^H_{12}-\de Z^H_{21}\bigr) 
       = \frac{\Sigma^H_{12}(M_{\PH_2}^2)+\Sigma^H_{12}(M_{\PH_1}^2)}{2(M_{\PH_1}^2-M_{\PH_2}^2)},
\end{align}
which is evidently symmetric and process independent.

The counterterm $\de\al$ appears in $S$-matrix elements only in the combinations
\begin{align}\label{eq:dzmda}
&{-\de\al} +\frac{1}{2}\de Z^H_{12}, \qquad
\de\al +\frac{1}{2}\de Z^H_{21} 
&& \mbox{from {the} field rotation } R(\alpha),\notag\\
&(M_{\PH_1}^2-M_{\PH_2}^2)\de\al
&& \mbox{from relations to parameters of {the} Higgs potential}.
\end{align}
With the condition \refeq{de_al_rigid}, the first two combinations become
\beq
\frac{1}{2}\de Z^H_{12}-\de\al = \de\al +\frac{1}{2}\de Z^H_{21} =
\frac{1}{4}\bigl(\de Z^H_{12}+\de Z^H_{21}\bigr)
 =
 \frac{\Sigma^H_{12}(M_{\PH_2}^2)-\Sigma^H_{12}(M_{\PH_1}^2)}{2(M_{\PH_1}^2-M_{\PH_2}^2)}.
\eeq
Thus, all three combinations \refeq{eq:dzmda} are smooth in the limit
of degenerate {scalar} masses {$M_{\PH_1}\to M_{\PH_2}$.}

Since the field renormalization constants $\de Z^H_{ij}$ are gauge
dependent, this {gauge dependence enters}
$\de\al$ defined in \refeq{de_al_rigid}.
However, as suggested in
\citeres{Kanemura:2015mxa,Yamada:2001px,Pilaftsis:2002nc} for the
renormalization of $\beta$, one can choose a specific gauge to
calculate the counterterms for $\de\al$ and thus fix this
renormalization constant once and for all. A convenient choice for
this purpose, which does not introduce large artificial parameters, is
the 't~Hooft--Feynman gauge.  Once $\de\al$ is kept fixed, $S$-matrix
elements are gauge-independent {functions of the renormalized input parameters}
as usual.

The method discussed can be applied to the renormalization of $\beta$
in the THDM by using the mixing between the physical pseudoscalar {Higgs boson} and
the would-be Goldstone boson. Alternatively, one can rely on the
definition $\tan\beta=\vtwo/\vone$ and the framework of the background
field method (BFM) \cite{Abbott:1980hw,Denner:1994xt,Denner:2017vms}.
{In the BFM,} fields are split into background fields $\Phihat$, which
serve as external sources, and quantum fields $\Phi$, which are
quantized. By choosing a suitable gauge-fixing term for the quantum
fields, rigid gauge invariance is maintained for the background fields
\cite{Denner:2017vms,Denner:1994xt}. This invariance implies Ward
identities and restrictions on renormalization constants for the
background fields. For instance, the relations \refeq{eq:Zsymrel} can
be kept exact including finite parts by choosing the field
renormalization constants appropriately.

Applying the BFM to the  THDM \cite{Denner:2017vms} yields the relations
for the symmetric field renormalization constants of the scalar fields,
\begin{align}
%\label{eq:rcrel2hdm1}
\delta Z^{\etahat}_1 
      = - 2 \delta Z_e - 
        \frac{\cw ^2}{\sw^2} \frac{\delta \cw ^2}{\cw ^2} + 
        \frac{\delta \MW^2}{\MW^2} +    2\frac{\Delta \vone}{\vone}
        + 2\frac{\delta\cbe}{\cbe},\notag\\ 
\label{eq:rcrel2hdm2}
\delta Z^{\etahat}_2 
      = - 2 \delta Z_e - 
        \frac{\cw ^2}{\sw^2} \frac{\delta \cw ^2}{\cw ^2} + 
        \frac{\delta \MW^2}{\MW^2} +    2\frac{\Delta \vtwo}{\vtwo}
        + 2\frac{\delta\sbe}{\sbe},
\end{align}
where the shifts $\Delta v_i$ of the vevs result from tadpole
contributions and can be expressed by tadpole counterterms ${\delta
  t}_{\hat H_i}$ as
\begin{align}
\Delta \vone = -\frac{\dtHhatone}{\MHsone}\ca + \frac{\dtHhattwo}{\MHstwo}\sa,\qquad
\Delta \vtwo = -\frac{\dtHhatone}{\MHsone}\sa - \frac{\dtHhattwo}{\MHstwo}\ca.
\end{align}
Using the relations \refeq{eq:Zsymrel} to eliminate $\delta
Z^{\etahat}_1$ in favour of $\delta Z^{\Hhat}_{ij}$ and solving for
$\delta\be$ yields \cite{Denner:2018opp}
\begin{align}
\label{eq:debeta_BFM_2HDM_new}
\delta\be={}&\frac{1}{2}\cbe\sbe\left[(\sa^2-\ca^2)\left(\de Z^{\Hhat}_{11}-\de Z^{\Hhat}_{22}\right)+2\ca\sa\left(\de Z^{\Hhat}_{12}+\de Z^{\Hhat}_{21}\right)\right]
%\notag\\&{}
+\frac{e}{2\sw\MW}\left(\sbe\Delta \vone -\cbe\Delta \vtwo\right).
\end{align}
This renormalization of the mixing angle $\beta$ is symmetric, process
{independent,} and smooth in the limits of degenerate Higgs masses and
extreme mixing angles. A similar renormalization condition {was}
proposed in \citere{Krause:2016oke}.

\section{Renormalization of mixing angles in the $\MSbar$ scheme}

{The $\MSbar$ scheme} offers a simple framework for the
renormalization of mixing angles. In this scheme, the renormalization
constants $\de\al$ contain only UV-divergent parts along with some
universal finite constants. The $\MSbar$ renormalization of mixing
angles is by construction symmetric in the mixing degrees of freedom
and independent of a specific observable. The counterterms are
{gauge independent} if the FJTS is used for the treatment of tadpoles.
{Finally, if perturbatively stable, the residual renormalization-scale 
dependence of calculated observables
offers a diagnostic tool for estimating residual theoretical uncertainties.}

The counterterm $\delta\al$ can for instance be fixed by taking the
UV-singular part of \refeq{de_al_rigid}. However, since the field
renormalization constants $\de Z^H_{ij}$ enter the $S$-matrix elements
including their complete finite part owing to the LSZ formalism, the
cancellation in the terms in the first line of \refeq{eq:dzmda}
becomes incomplete, and these contributions {become} singular for
$M_{\PH_i}\to M_{\PH_j}$. As a consequence, all $\MSbar$
renormalization schemes for mixing angles give rise to large
corrections in the limit of degenerate masses.  These effects are
enhanced by additional tadpole contributions in the FJTS.

\section{Renormalization schemes for mixing angles based on 
{\boldmath{$S$-matrix elements}}}

Defining the mixing angles from single observables {in general} leads to
unnaturally large corrections \cite{Krause:2016oke}, since these
observables depend on further parameters. A better strategy uses
combinations of observables 
{or $S$-matrix elements} that depend exclusively on a mixing angle.
Consider as an example the LO matrix elements for the decay of the
scalar Higgs bosons $H_1$ and $H_2$ into a pair of \PZ~bosons in the
HSESM,
\begin{align}
\cM_0^{H_1\to \PZ\PZ} = \frac{e \sa}{\sw \cw^2} {\MW} (\varepsilon_1^*\cdot\varepsilon_2^*), \qquad
\cM_0^{H_2\to \PZ\PZ} = \frac{e \ca}{\sw \cw^2} {\MW} (\varepsilon_1^*\cdot\varepsilon_2^*),
\end{align}
where $\varepsilon_{1,2}$ denote the polarization vectors of the two Z~bosons.
As renormalization condition we can require that the ratio of these
two matrix elements, which at LO is a function of $\alpha$ only, does
not change under renormalization, \ie
\beq\label{eq:RC_phys_HZZ}
\frac{\cM^{H_1\to \PZ\PZ}}{\cM^{H_2\to \PZ\PZ}}  \overset{!}{=}
\frac{\cM_0^{H_1\to \PZ\PZ}}{\cM_0^{H_2\to \PZ\PZ}} = \frac{\sa}{\ca} =f(\al)
.
\eeq
In the complete on-shell scheme, this yields for the counterterm of
the mixing angle $\alpha$ at NLO
\begin{align}
\de\al={}& \ca\sa(\delta_{H_2ZZ}-\delta_{H_1ZZ}) +\frac{1}{2}\ca\sa(\delta Z^H_{22}-\delta Z^H_{11})
+\frac{1}{2}(\delta Z^H_{12}\satwo- \delta Z^H_{21}\catwo),
\label{eq:da_HZZ}
\end{align}
where $\de_{H_iZZ}=\de_{H_iZZ}(M_{\PH_i}^2)$ are the unrenormalized
relative one-loop corrections to the respective decay {matrix elements.}  This
renormalization condition is gauge independent, because it is based on
physical $S$-matrix elements; it is symmetric with respect to the
scalar fields $H_1$ and $H_2$; it is numerically stable for degenerate
masses $M_{\PH_1}\sim M_{\PH_2}$; and it has smooth limits for extreme
mixing angles, \ie for $\ca\to0$ or $\sa\to0$.  However, it is
restricted to processes with only neutral external particles, since
for charged particles the renormalization constant $\delta\al$
becomes IR singular, and for the observed Higgs boson with mass
$125\GeV$ the relative correction factors for the corresponding decay
have to be evaluated in the unphysical region.

These drawbacks can be lifted by introducing suitable extra neutral
fields with a simple coupling structure and considering the limit of
vanishing extra {couplings~\cite{Denner:2018opp}.}  As an example we
add an additional fermion {singlet~$\psi$} to the HSESM with the
Lagrangian% 
\footnote{{We choose a fermion singlet in order to limit
    the number of independent operators that could mix when
    renormalizing the extended model.}}
\beq
\cL_\psi = \ri\bar\psi\slashed{\partial}\psi - y_\psi \sigma\bar\psi\psi
= \ri\bar\psi\slashed{\partial}\psi - y_\psi(\vone+H_1\ca-H_2\sa) \bar\psi\psi.
\eeq
and consider the limit of vanishing Yukawa coupling $y_\psi$.
Requiring that the ratio of the matrix elements for the decays of the
two scalar Higgs bosons into a $\bar\psi\psi$ pair does not receive
any higher-order corrections in the limit of vanishing $y_\psi$,
\beq\label{eq:RC_phys_Hchichi}
\frac{\cM^{H_1\to \psi\psi}}{\cM^{H_2\to \psi\psi}}
\overset{!}{=} 
\frac{\cM_0^{H_1\to \psi\psi}}{\cM_0^{H_2\to \psi\psi}}
\propto -\frac{\ca}{\sa},
\eeq
fixes the {mixing-angle counterterm} as
\begin{align}
\de\al={}& \frac{1}{2}(\delta Z^H_{11}-\delta Z^H_{22}) \ca\sa
+\frac{1}{2}(\delta Z^H_{12}\catwo-\delta Z^H_{21}\satwo).
\label{eq:da_hsesm_os}
\end{align}
Owing to the simple structure of the model the vertex corrections drop
{out,} and the counterterm is fixed by a (gauge-independent) combination
of field-renormalization constants only.

This strategy can be applied to the THDM as well. {To this end,} we add
two right-handed fermion singlets to the Lagrangian,$\nu_{1\rR}$
transforming under the extra $\mathbb{Z}_2$ symmetry as
$\nu_{1\rR}\to-\nu_{1\rR}$ and $\nu_{2\rR}$ transforming as
$\nu_{2\rR}\to\nu_{2\rR}$, so that $\nu_{i\rR}$ can only receive a
Yukawa coupling to $\Phi_i$. The additional Lagrangian is given by 
\begin{align}
\cL_{\nu_{\rR}} ={}& 
\ri\bar\nu_{1\rR}\slashed{\partial}\nu_{1\rR}
+\ri\bar\nu_{2\rR}\slashed{\partial}\nu_{2\rR}
-\left[ y_{\nu_1} \bar{L}_{1\rL} (\ri\sigma_2\Phi^*_1)\nu_{1\rR} 
+ y_{\nu_2} \bar{L}_{2\rL} (\ri\sigma_2\Phi^*_2)\nu_{2\rR} +\hc  \right], 
\end{align}
and the new Yukawa couplings are considered in the limit
$y_{\nu_i}\to0$.
Renormalization conditions can be formulated \cite{Denner:2018opp} by
requiring that ratios of {$S$-matrix} elements or relations of form
factors of Higgs decays into fermion singlets do not get
perturbative corrections. Thus, one arrives at the following
renormalization constants for the mixing angles, 
%of the THDM:
\begin{align}
\de\al={}&(\delta_{H_1\nunuone}-\delta_{H_2\nunuone}) \ca\sa 
+\frac{1}{2}(\delta Z^H_{11}-\delta Z^H_{22})\ca\sa
+\frac{1}{2}(\delta Z^H_{12}\catwo- \delta Z^H_{21}\satwo),
 \label{eq:da_nu1_ons} \\
 \de\beta={}&\frac{1}{2}\cbe\sbe\left[(\ca^2-\sa^2)(\de Z^H_{11}-\de
   Z^H_{22}) -2\ca\sa(\de Z^H_{12}+\de Z^H_{21})\right]
+ \frac{1}{2}\de Z_{G_0\Ha}
\notag\\&{}+
\cbe\sbe\left(
\delta_{\Ha\nunutwo}+\ca^2\de_{H_1\nunuone}+\sa^2\de_{H_2\nunuone}
-\delta_{\Ha\nunuone}-\sa^2\de_{H_1\nunutwo}-\ca^2\de_{H_2\nunutwo}
\right),
\label{eq:db_os12_thdm}
\end{align}
where again
$\de_{H_i\nu_j\bar\nu_j}=\de_{H_i\nu_j\bar\nu_j}(M_{\PH_i}^2)$ are the
unrenormalized relative one-loop corrections to the respective decays.
These renormalization conditions are gauge independent, symmetric with
respect to the scalar fields $H_1$ and $H_2$ and behave smoothly in
the limits of degenerate masses $M_{\PH_1}\sim M_{\PH_2}$ and for
extreme mixing angles ($\sa\to0$, $\ca\to0$, $\sbe\to0$, or
$\cbe\to0$).

\section{Numerical results}

\begin{figure}
\centerline{
\ifpdf
\includegraphics[width=0.49\textwidth]{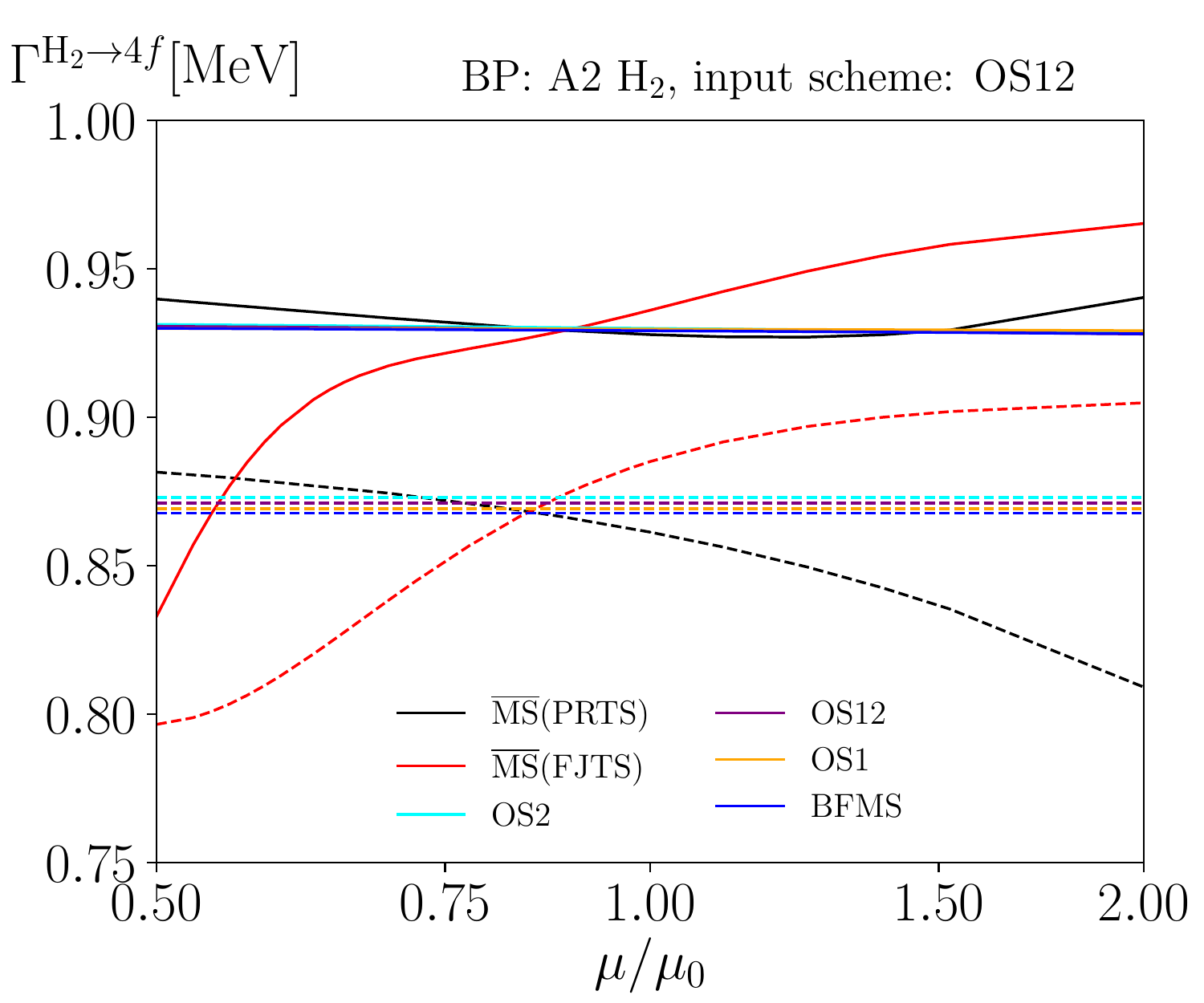}
\includegraphics[width=0.49\textwidth]{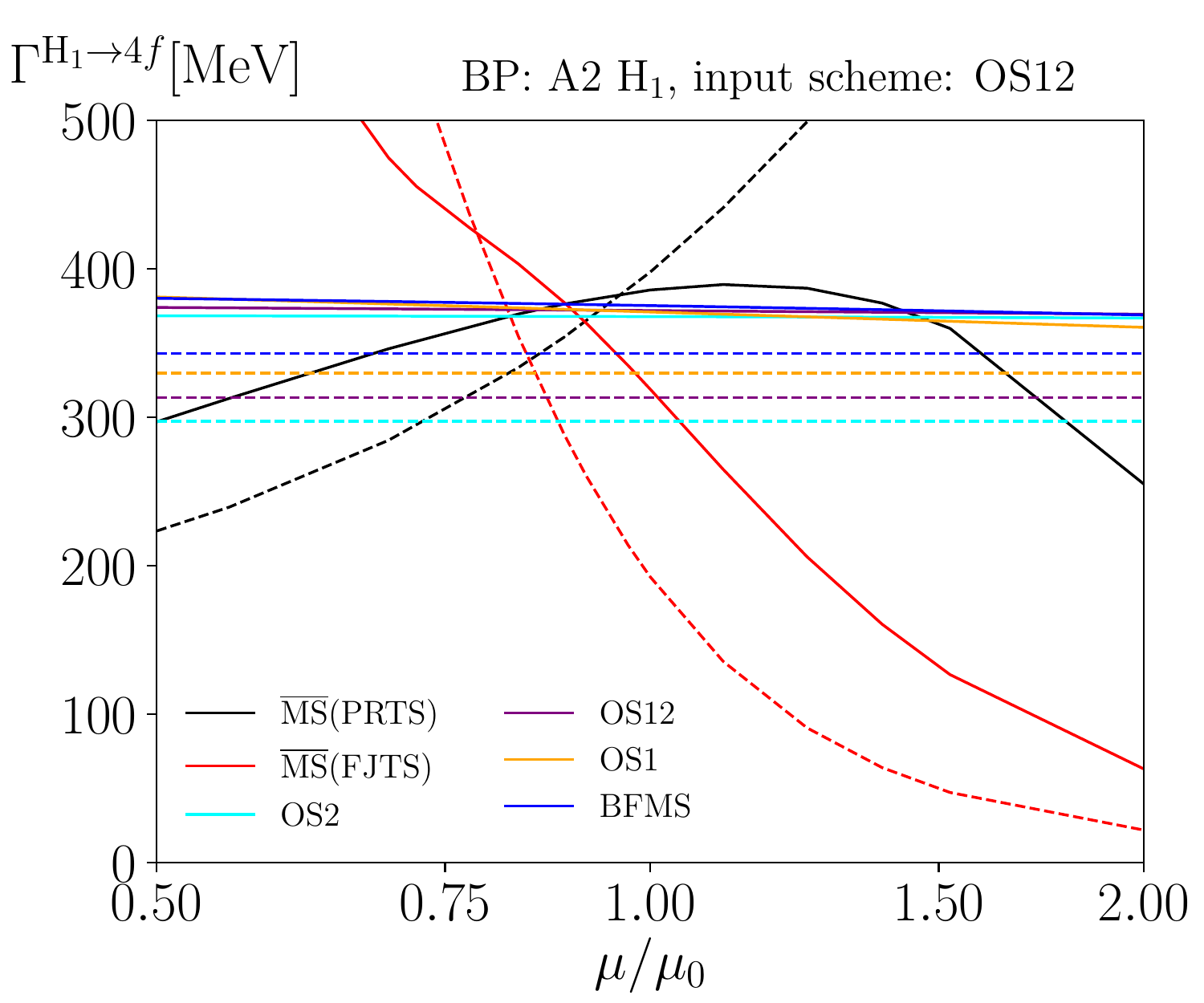}
\else
\includegraphics[width=0.49\textwidth]{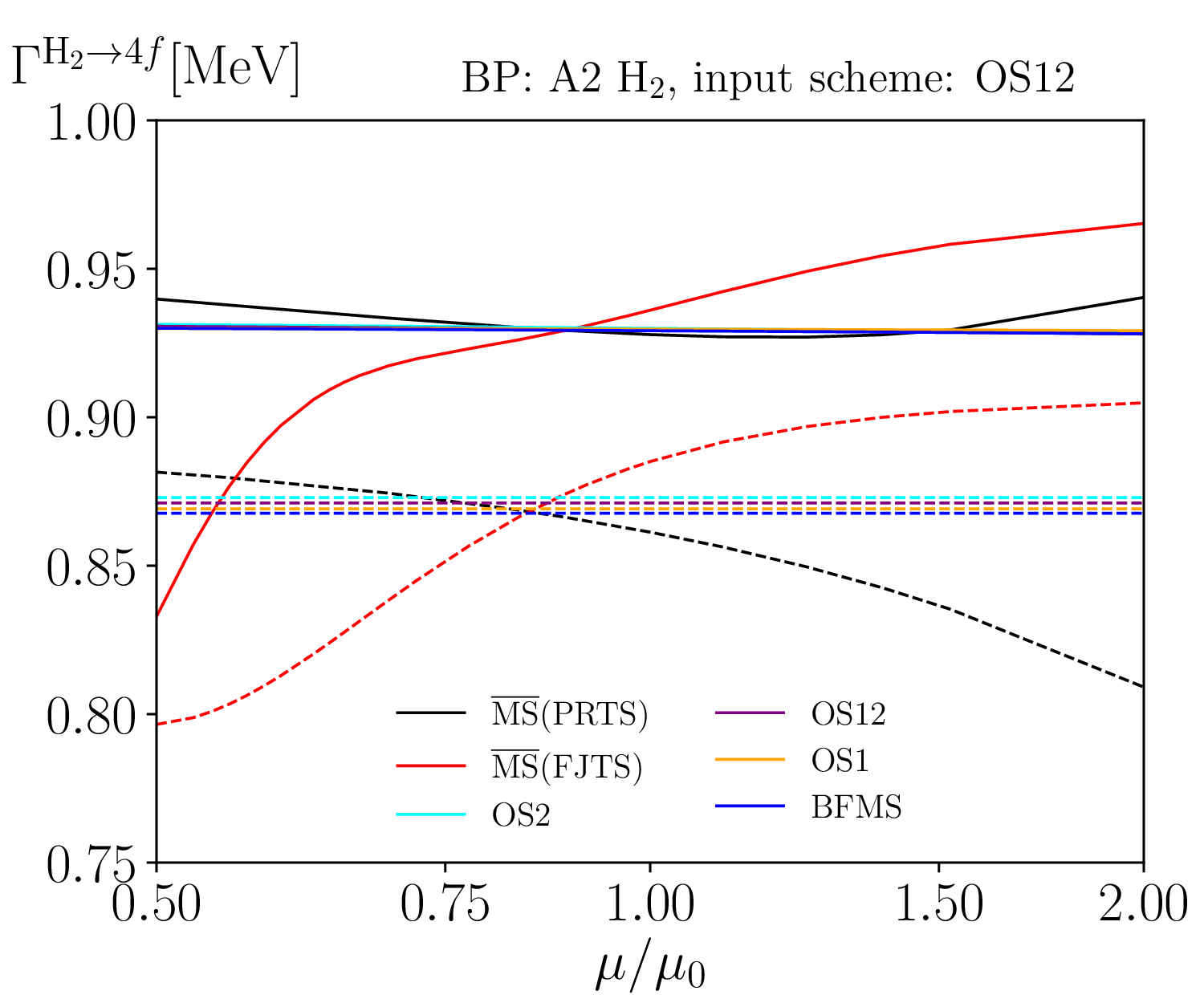}
\includegraphics[width=0.49\textwidth]{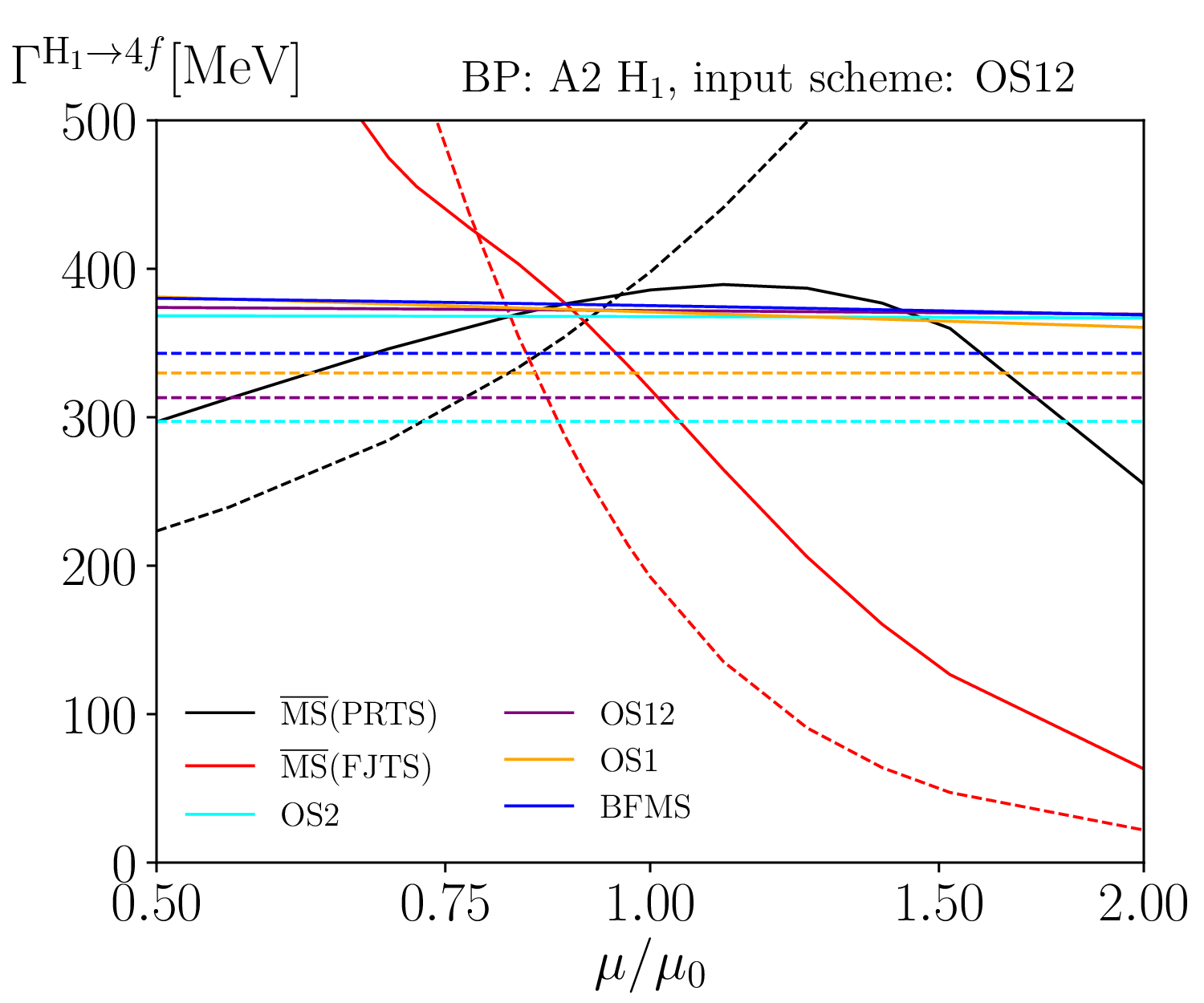}
\fi
}
\caption{Scale dependence of the decay widths 
  of the light (left) and heavy (right) \THDM\ Higgs bosons for the
  \THDM\ scenario A2 in different renormalization schemes, with the
  \OSonetwo\ scheme as input scheme.  LO results are shown as dashed,
  NLO as full lines; the central scale is set to
  $\mu_0=(M_{\PH_2}+M_{\PH_1}+M_{\Ha}+2M_{\PH^+})/5$.
  {(Plots taken from \citere{Denner:2018opp}.)}}
\label{fig:THDM-H4f_A2}
\end{figure}

For illustration, in \reffi{fig:THDM-H4f_A2} we provide numerical
results for the decay widths of the light and heavy scalar Higgs
bosons into four fermions in a specific scenario of the THDM obtained
with an extended version of
\Prophecy~\cite{Bredenstein:2006rh,Bredenstein:2006ha,Denner:2019fcr}.
The renormalization schemes as well as the input parameters are
defined in \citere{Denner:2018opp}. The input parameters are fixed
in the scheme OS12, where the renormalization of the mixing angles is
based on \refeq{eq:da_nu1_ons} and \refeq{eq:db_os12_thdm}, and
converted with NLO precision to the other schemes.

The plots show the scale dependence of the decay widths in the
$\MSbar$ renormalization schemes for the mixing angles $\alpha$ and
$\beta$ in the two tadpole schemes PRTS and FJTS, 
{as calculated in \citeres{Altenkamp:2017ldc,Altenkamp:2017kxk}.}
While the scale
dependence of the decay width of the light Higgs boson is reduced to
2\% at NLO in the PRTS, no reduction is found in the FJTS.  For the
decay of the heavy Higgs boson neither of the $\MSbar$ schemes
provides a reduction of scale dependence.
{In general, the perturbative stability of the results obtained with
$\MSbar$-renormalized mixing angles $\alpha$ and $\beta$ strongly depends 
on the considered scenario (see 
\citeres{Altenkamp:2017ldc,Altenkamp:2017kxk,Denner:2018opp} for more
examples).}

The schemes OS12, OS1, and OS2 (the latter two being variants of OS12)
based on sets of physical {$S$-matrix elements} and the scheme BFMS defined in
\refeq{de_al_rigid} and \refeq{eq:debeta_BFM_2HDM_new} nicely agree at
NLO. In general, uncertainty estimates should be based on schemes that
are well behaved and not on the scale dependence of 
{$\MSbar$ schemes if those turn out to be perturbatively unstable.}

Further numerical results in different scenarios for the Higgs decays
into four fermions and Higgs production in Higgs strahlung and
vector-boson fusion in the THDM and the HSESM can be found in
\citeres{Denner:2016etu,Denner:2017vms,Altenkamp:2017ldc,%
Altenkamp:2018bcs,Denner:2018opp,Altenkamp:2017kxk}. Results for
Higgs decays into 2-particle final states in the THDM were presented
in \citere{Krause:2016oke}. 

The renormalization schemes based on rigid symmetry and BFM as well as
the renormalization schemes based on combinations of observables have
been implemented in {\sc 2HDECAY} \cite{Krause:2018wmo}, \Prophecy 3.0
\cite{Denner:2019fcr}, and {\sc HAWK 3.0} \cite{Denner:2017wsf}.

\section{Conclusions}
The mechanism of electroweak symmetry breaking and specifically models with
extended Higgs sectors are studied with high accuracy at the LHC. This
requires to calculate higher-order corrections in these models which
in turn needs renormalization of their parameters including mixing angles.
In this contribution we have presented renormalization schemes for
mixing angles in the THDM and HSESM that exhibit symmetry with respect
to the mixing degrees of freedom, {gauge independence,} perturbative
stability as well as applicability in the full parameter space.
The schemes are defined via genuine on-shell renormalization
conditions or symmetry principles, 
{so that their generalization to
higher orders is well defined and their application to
other extensions of the Standard Model {feasible}.
With the implementation of the various types and variants of 
renormalization schemes in the programs {\sc 2HDECAY}, 
\Prophecy~3.0 and {\sc HAWK 3.0}, significant steps towards
precision Higgs analysis in the THDM and HSESM are made on the theory side.}

\acknowledgments The work of A.D.\ is supported by the German Science
Foundation (DFG) under reference number DE~623/5-1, 
the work of S.D.\ by the DFG project DI~784/4-1, by the state of
Baden-W\"urttemberg through bwHPC and the DFG through grant no INST
39/963-1 FUGG.  
{J.-N.~L.\ acknowledges} support from the Swiss National Science
Foundation (SNF) under contract BSCGI0-157722.

%\begin{thebibliography}{99}
%\bibitem{...}
%....
%\end{thebibliography}

\bibliographystyle{JHEPmod}
\bibliography{mixing}

\end{document}